\documentclass{ws-ijmpa}

\begin{document}

\markboth{H. Arakida and S. Kuramata}
{Diffusive Propagation of High Energy Cosmic Rays}

%
\catchline{}{}{}{}{}
%

\title{
Diffusive Propagation of\\
High Energy Cosmic Rays in Galaxy:\\
Effect of Hall Drift
}

\author{Hideyoshi Arakida}

\address{School of Education, Waseda University
\\
arakida@edu.waseda.ac.jp}

\author{Shuichi Kuramata}

\address{
Graduate School of Science and Technology, Hirosaki University
}

\maketitle

\begin{history}
\received{Day Month Year}
\revised{Day Month Year}
\end{history}

\begin{abstract}
We phenomenologically developed a propagation model of high energy 
galactic cosmic rays. We derived the analytical 
solutions by adopting the semi-empirical diffusion equation,
proposed by Berezinskii {\it et al.}(1990) and the diffusion 
tensor proposed by Ptuskin {\it et al.}(1993). This model
takes into account both the symmetric diffusion and
the antisymmetric diffusion due to the particle 
Hall drift. Our solutions are an extension of the model 
developed by Ptuskin {\it et al.} (1993) to a two-dimensional 
two-layer (galactic disk and halo) model, and they coincide 
completely with the solution derived by Berezinskii {\it et al.} 
(1990) in the absence of antisymmetric diffusion  due to Hall drift. 
We showed that this relatively simple toy model can be used 
to explain the variation in the exponent of the cosmic ray 
energy spectrum, $\gamma$, around the knee $E \approx 10^{15}$ eV.
\end{abstract}

\ccode{PACS numbers: 96.50.S-, 98.70.Sa, 96.50.sb}

\section{Introduction\label{intro}}
Cosmic ray propagation is one of the most important and interesting
subjects in astrophysics and
high energy particle physics. It is believed that the observed 
cosmic ray data such as the cosmic ray energy spectrum
includes information about the space through which the cosmic 
rays pass. In fact, it is possible to evaluate the thickness of 
matter, and it is also thought that details about the galactic 
magnetic field can be extracted from the observed data because
the cosmic rays experience frequent collision and scattering with both the 
interstellar gas and the galactic magnetic field during 
their propagation in the galaxy. Thus, a reliable cosmic 
ray propagation model may enable us to obtain further knowledge 
about the galactic structure. Thus far, several cosmic ray 
propagation models have been proposed and discussed 
(see \cite{cesarsky,ptuskin2001} and references therein).
Further, some researchers have successfully derived analytical 
solutions using the diffusion equation
\cite{berezinskii,pacheco,guet,ptuskin}.

The spectrum of cosmic rays shows one of the most distinctive 
features, known as ``knee'', around 
energy $E \approx 10^{15}$ eV at which the exponent of the 
energy spectrum, $\gamma$, changes from 2.6 -- 2.7 for 
$10^{10} \le E \le 10^{15}$ eV to 3 -- 3.1 for 
$10^{15} \le E \le 10^{18}$ eV. Currently, it is not clear 
why the exponent changes around $E \approx 10^{15}$ eV. 
Thus far, several models have been proposed to explain this spectral 
property: a shock wave acceleration model based on the acceleration 
of cosmic ray particles by the shock wave front
\cite{berezhko,stanev,kobayakawa,sveshnikova,erlykin,volk,plaga}, 
a diffusive propagation model based on the leakage and diffusive 
propagation of cosmic rays in the galaxy 
\cite{swordy,lagutin,ptuskin,ogio,roulet,candia1,candia2,candia4}.
an interaction model based on the interaction of cosmic rays 
with the background particles in the galaxy 
\cite{tkaczyk,karakula,dova,candia3},
a reaction model based on the reaction of cosmic rays with 
the atmosphere of Earth \cite{kazanas1,kazanas2}, etc.
Among these models, the shock wave acceleration model seems to be 
widely accepted as the explanation for the knee. 
However, the diffusive propagation model can provide the 
exposition for the first knee and the second knee and 
for the observed compositions and 
anisotropies \cite{ptuskin,candia1,candia2,candia4}. 
The diffusive propagation model is characterized by 
introducing the particle Hall drift effect; thus, 
it is a theoretically simple model. 

In this study, we derived solutions for the diffusive propagation 
model of cosmic rays and confirmed the validity of this model. 
We adopted the semi-empirical diffusion equation introduced 
in \cite{berezinskii} and the diffusion tensor described in
\cite{ptuskin}. Then, we extended the propagation model in
\cite{ptuskin} to a two-dimensional two-layer (comprising the 
galactic disk and the halo) model in cylindrical coordinates.

This paper is organized as follows: In Section \ref{solution}, we
derive the analytical solutions for the diffusion equation of 
cosmic rays. In Section \ref{psols}, we qualitatively show that
our model can explain the spectral feature of the observed cosmic 
rays, namely, the exponential variation around the knee. In Section 
\ref{conclusion}, we present the conclusions of our study.
\section{Solutions for Diffusion Equation\label{solution}}
\subsection{Galactic Structure}
\begin{figure}
 \begin{center}
  \includegraphics[scale=0.5]{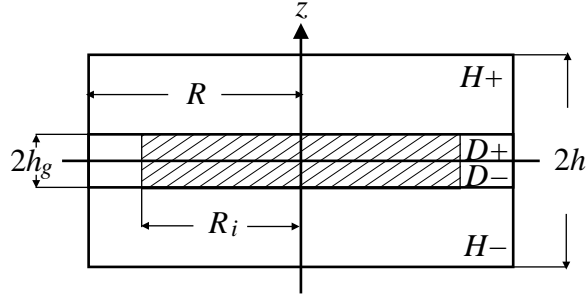}
  \caption{Schematic Diagram of Galactic Structure. \label{gstructure}}
\end{center}
\end{figure}

Fig. \ref{gstructure} shows a schematic diagram of our 
galactic model. We assume that our galaxy has a cylindrical 
structure and it is divided into two parts; the galactic disk ($D$) 
and the halo ($H$). The signs $+$ and $-$ indicate the direction 
along the $z$-axis. In Fig. \ref{gstructure}, $R$ is the radius of 
the galaxy and $2h_g$ and $2h$ are the height of the disk 
region and the galaxy, respectively. 
We suppose that the source of cosmic rays is 
distributed uniformly within the galactic disk area only. 
The source region of the $i$-th particle is given by 
radius $R_i$ ($0 < R_i \le R$) in the shaded area in Fig. 
\ref{gstructure}.
\subsection{Basic Equation, Diffusion Tensor, and Hall Drift}
We consider the following transport equation
\cite{berezinskii}
\begin{equation}
  - \nabla_{\alpha} (D_{\alpha \beta}(z) \nabla_{\beta})N_{i}(E, r, z)
  + n(z)v \sigma_{i}N_{i}(E, r, z) = Q_{i}(E, r, z),
\label{diff-eq2}
\end{equation}
where $N_i(E, r, z)$ is the number density of the $i$-th 
cosmic ray component per unit time, unit volume, and unit energy; 
$D_{\alpha \beta}$ is the diffusion tensor (indexes 
$\alpha$ and $\beta$ denote the coordinates); $v$ is the 
velocity of the cosmic ray; $\sigma_i$ is the inelastic 
scattering cross-section between the $i$-th cosmic ray particle 
and the interstellar gas; $n(z)$ is the density of the interstellar 
gas; and $Q_i(E, r, z)$ represents the distribution of the source. 

We adopt the diffusion tensor given in \cite{ptuskin} 
\begin{equation}
  D_{\alpha \beta} = (D_{\Vert} - D_{\bot})b_{\alpha}b_{\beta}
  + D_{\bot}\delta_{\alpha \beta} + D_{A} \epsilon_{\alpha \beta \gamma}
  b_{\gamma},
  \label{tensor}
\end{equation}
where $b_{\alpha}$ is the unit vector of the magnetic field,
$D_{\Vert}$ denotes the diffusion along the magnetic
field line, $D_{\bot}$ denotes the diffusion
perpendicular to the magnetic field line,
$D_{A}$ is diffusion due to the antisymmetric drift of the particle,
$\delta_{\alpha\beta}$ is the Kronecker's delta symbol, and 
$\epsilon_{\alpha \beta \gamma}$ is the complete 
antisymmetric Levi-Civita tensor. 

The observed data show that the magnetic field in the 
galaxy has a spiral structure, and it is regarded 
as an approximate toroidal structure \cite{heiles}. Thus,
we assume the unit vector of the magnetic field as
\begin{eqnarray}
 b_{r} = b_{z} = 0,\quad b_{\phi} = 1.
\label{uvector1}
\end{eqnarray}
It should be noted that the magnetic field line in the galactic 
disk changes the sign almost every 3 kpc along the radial 
direction \cite{rand}. Nevertheless, for the sake of simplicity,
we assume that the global magnetic field points in the same 
direction anywhere in the galactic disk. 
On the other hand, the structure of the magnetic field in 
the galactic halo is not well known; so we suppose that the magnetic
field in the halo also has a toroidal structure given by 
\begin{eqnarray}
 b_{r}^{(D)} = b_{r}^{(H)} = b_{z}^{(D)} = b_{z}^{(H)} = 0,\quad
 b_{\phi}^{(D)} = b_{\phi}^{(H)} = 1,
\label{uvector2}
\end{eqnarray}
where $(D)$ and $(H)$ denote the disk and the halo, respectively.
From Eq. (\ref{uvector2}), the diffusion terms in 
(\ref{diff-eq2}) becomes
\begin{equation}
  -\nabla_{\alpha}(D_{\alpha \beta} \nabla_{\beta})
  = -\frac{1}{r}\frac{\partial}{\partial r}(r D_{\bot})
  \frac{\partial}{\partial r}
  - \frac{\partial}{\partial z}D_{\bot}\frac{\partial}{\partial z}
  - \frac{\partial}{\partial z}D_{A}\frac{\partial}{\partial r}
  + \frac{1}{r}\frac{\partial}{\partial r}(rD_{A})
  \frac{\partial}{\partial z}.
  \label{dtensor1}
\end{equation}
These coefficients are determined in
\footnote{
It is suggested that the turbulent exponent in the magnetic 
field in the galactic disk is practically characterized by 
the power law with exponent $m = 0.3 \pm 0.3$ 
\cite{armstrong,ruzmaikin}. For instance, $m=1/3$ is related to 
the Kolmogolov turbulence spectrum; $m = 1/2$,  
turbulence spectrum of Kraichinan hydromagnetic one 
\cite{kraichinan}; and $m = 0$, Bykov-Toptygin 
spectrum \cite{bykov}. In the case of the halo, we do not have 
sufficient knowledge about the magnetic turbulence;
therefore, we assume that the same power law holds} as
\begin{eqnarray}
 D_{\Vert} = \frac{lv}{3},\quad 
  D_{\bot} = gA^4\frac{lv}{3},\quad
  D_{A} = - \frac{r_{H_0}v}{3}.
\end{eqnarray}
Here $l$ is the mean free path that is characterized by the
turbulent exponent $m$, $A$ denotes the relative value of a 
random magnetic field within the characteristic scale $L$, 
and $g$ is a not well-determined parameter (see \cite{ptuskin}).
From the gyro-radius
\begin{eqnarray}
 \mbox{\boldmath $r$}_{\mbox{\boldmath $H$}_0} = \frac{c}{Ze}
\frac{\mbox{\boldmath $B$}\times\mbox{\boldmath $p$}}{B^2},
\end{eqnarray}
we obtain the drift velocity of the particles
\begin{eqnarray}
 V_{D\gamma} = -\nabla_{\gamma}(D_A\epsilon_{\alpha\beta\gamma}b_{\alpha})
  = \frac{pcv}{3Ze}\epsilon_{\alpha\beta\gamma}
  \frac{\partial}{\partial x_{\alpha}}\frac{B_{\beta}}{B^2}.
\end{eqnarray}
Then, we substitute the coefficients of the diffusion tensor 
given in \cite{ptuskin} 
\begin{eqnarray}
    D_{\bot}(r,z) = \mbox{constant},\quad 
    D_{A}(r,z) = \displaystyle{D_{A0}(z)\frac{r}{R}},\quad
    D_{A0}(z) = \mbox{constant}.
  \label{coeff}
\end{eqnarray}
Then the diffusion term reduces to
\begin{equation}
  -\nabla_{\alpha}(D_{\alpha \beta} \nabla_{\beta})
  =\frac{D_{\bot}}{r}\frac{\partial}{\partial r} r \frac{\partial}{\partial r}
  - D_{\bot}\frac{\partial^{2}}{\partial z^{2}}
  + \frac{2 D_{A0}}{R} \frac{\partial}{\partial z}.
  \label{dtensor2}
\end{equation}
From Eq. (\ref{coeff}), the drift velocity $V_D$ is written as
\begin{eqnarray}
  V_{Dr} = -\frac{\partial}{\partial z}D_A = 0,\quad 
  V_{Dz} = \frac{1}{r}\frac{\partial}{\partial r}(rD_A) 
  = \frac{2D_{A0}}{R}  = \mbox{constant}.
\end{eqnarray}
Thus, the drift effect appears only along the 
$z$-direction in this case.
\subsection{Boundary and Continuity Conditions}
Let us summarize the boundary and continuity conditions. 
First, at the edge of the galaxy, we assume
that the cosmic rays leak from the galaxy
\begin{eqnarray}
  N_i(r = R, z) = N_i(r, z = \pm h) = 0.
   \label{boundary1}
\end{eqnarray}
This condition seems to be appropriate in terms of the cosmic ray 
$L/M$ ratio. Second, at the boundaries of the disk and the halo, 
$z = \pm h_g$, both $N_i^{(D)}$ and $N_i^{(H)}$ must be 
continuously connected. Then,
\begin{eqnarray}
 N_{i}^{(H\pm)}(\pm h_{g}, r) = N_{i}^{(D\pm)}(\pm h_{g}, r),\quad
  \frac{d N_{i}^{(H\pm)}(\pm h_{g}, r)}{d z} =
  \frac{d N_{i}^{(D\pm)}(\pm h_{g}, r)}{d z},
  \label{boundary2}
\end{eqnarray}
and 
\begin{eqnarray}
N_{i}^{(D+)}(0, r) = N_{i}^{(D-)}(0, r),\quad
\frac{dN_{i}^{(D+)}(0, r)}{d z}
= \frac{dN_{i}^{(D-)}(0, r)}{d z}.
\label{boundary3}
\end{eqnarray}
\subsection{Solution for Transport Equation\label{solutionofeq}}
Because Eq. (\ref{diff-eq2}) has the same form as Eqs. 
(3.10) or (3.11) in \cite{berezinskii}, we follow the same approach.
To obtain the solution for Eq. (\ref{diff-eq2}), we first obtain the 
Green's function $\Phi(r, z; r_{0})$, which satisfies 
\begin{equation}
  -\nabla_{\alpha}(D_{\alpha \beta}(z) \nabla_{\beta})\Phi(r, z; r_{0})
  + n(z)v\sigma_{i}\Phi(r, z; r_{0})
  = \theta(h_{g} - |z|)\frac{\delta(r - r_{0})}{4 \pi h_{g} r_{0}},
\label{green}
\end{equation}
where $r_{0}$ represents the position of the cosmic ray source
and $\theta(h_{g} - |z|)$ is the step function. $\Phi(r, z; r_{0})$ 
is related to $N_i$ in \cite{berezinskii}:
\begin{equation}
  N_{i} 
  = 4 \pi h_{g} g_{i} E^{-\gamma_{i0}}
  \int_{0}^{R_i} \chi(r_{0}) \Phi(r, z; r_{0}) r_{0}dr_{0},
\label{nphi}
\end{equation}
where $g_i$ is the constant characterizing the $i$-th 
particle, $\chi(r_{0})$ represents the radial distribution of
the source of cosmic rays, and we presume the energy spectrum at 
the source obeys the power law $E^{-\gamma_{i0}}$ for the 
$i$-th particle. We also suppose the radial component of 
$\Phi(r, z; r_{0})$ is expanded by using the zeroth-order 
Bessel function
\begin{equation}
  \Phi(r, z; r_{0})
  = \sum_{s = 1}^{\infty}a_{s}(z ; r_{0})
  J_{0} \left(\frac{\nu_{s} r}{R}\right).
\end{equation}
In this expression, $\nu_{s}$ is the $s$-th root of the equation
$J_{0}(\nu_{s}) = 0$. Hence, Eq. (\ref{green}) becomes
\begin{eqnarray}
  & &
  \sum_{s = 1}^{\infty}
  \left\{
    \left[
      -\frac{D_{\bot}(z)}{r}\frac{\partial}{\partial r} 
      r \frac{\partial}{\partial r}
      - D_{\bot}(z)\frac{\partial^2}{\partial z^2}
      + \frac{2D_{A0}(z)}{R}\frac{\partial}{\partial z}
    \right]
    + n(z)v\sigma_{i}
  \right\}a_{s}(z; r_{0})J_{0}\left(\frac{\nu_{s} r}{R}\right)
  \nonumber\\
  &=& \theta(h_{g} - |z|)\frac{\delta(r - r_{0})}{4 \pi h_{g} r_{0}}.
  \label{green2}
\end{eqnarray}
From the orthogonality of the Bessel function, 
the delta function is expressed as
\begin{equation}
\delta(r - r_{0}) = 
\sum_{s = 1}^{\infty}\frac{2}{|J_{1}(\nu_{s})|^2}
\frac{r_0}{R^2}
J_{0}\left(\frac{\nu_{s}r_{0}}{R}\right)
J_{0}\left(\frac{\nu_{s}r}{R}\right),
\end{equation}
where $J_{1}$ is the first-order Bessel function. 
Noting that
\begin{equation}
\frac{1}{r}\frac{\partial}{\partial r}r\frac{\partial}{\partial r}
J_{0}\left(\frac{\nu_{s}r}{R}\right)
=
-\left(\frac{\nu_{s}}{R}\right)^2 
J_{0}\left(\frac{\nu_{s}r}{R}\right),
\end{equation}
Eq. (\ref{green2}) is written as
\begin{eqnarray}
  & &
  \sum_{s = 1}^{\infty}
  \left\{
    \left[
      D_{\bot}(z)\left(\frac{\nu_{s}}{R}\right)^2
      - D_{\bot}(z)\frac{\partial^2}{\partial z^2}
      + \frac{2 D_{A0}(z)}{R}\frac{\partial}{\partial z}
    \right]
    + n(z)v\sigma_{i}
  \right\}
  a_{s}(z; r_{0})J_{0}\left(\frac{\nu_{s} r}{R}\right) 
  \nonumber\\ 
  &=& 
  \sum_{s = 1}^{\infty}
  \frac{\theta(h_{g} - |z|)}{2 \pi h_{g} R^2 |J_{1}(\nu_{s})|^2}
  J_{0}\left(\frac{\nu_{s}r_{0}}{R}\right)
  J_{0}\left(\frac{\nu_{s}r}{R}\right). 
\end{eqnarray}
Thus, the problem reduces to solving the ordinary differential 
equation
\begin{eqnarray}
  \left[
    \frac{d^2}{d z^2}
    -\frac{2 D_{A0}(z)}{R D_{\bot}(z)}\frac{d}{dz}
  \right]
  a_{s}(z; r_{0})
  -\left[
    \left(\frac{\nu_{s}}{R}\right)^2
    + \frac{n(z)v\sigma_{i}}{D_{\bot}(z)}
  \right]a_{s}(z; r_{0}) = - C,
\end{eqnarray}
where 
\begin{eqnarray}
 C = \frac{\theta(h_{g} - |z|)}
  {2 \pi h_{g} R^2 D(z)_{\bot}|J_{1}(\nu_{s})|^2}
  J_{0}\left(\frac{\nu_{s} r_{0}}{R}\right).
\end{eqnarray}
$D_{\bot}$, $D_{A0}$, and $n(z)$ are chosen as
\begin{eqnarray}
 D_{\bot}(z) =
 \left\{
 \begin{array}{r}
  D_{\bot}\\
  d_{\bot}\\
 \end{array}
\right.\quad
D_{A0}(z) = 
 \left\{
 \begin{array}{r}
  D_{A0}\\
  d_{A0}\\ 
 \end{array}
\right.\quad
n(z) = 
\left\{
  \begin{array}{rl}
    n_{g} & |z| \leq h_{g} \mbox{(disk)}, \\
    n_{h} & |z| > h_{g} \mbox{(halo)}. \\
  \end{array}
\right.
\end{eqnarray}
In the galactic halo, $C = 0$ because of the step function. 
Therefore,
\begin{eqnarray}
  \frac{d^2 a_{s}^{(H)}}{d z^2}
  - \frac{2 d_{A0}}{R d_{\bot}} \frac{d a_{s}^{(H)}}{d z}
  - \left[
    \left(\frac{\nu_{s}}{R}\right)^2
    + \frac{n_{h}v\sigma_{i}}{d_{\bot}}
  \right] a_{s}^{(H)}
  = 0,
  \label{ahalo}
\end{eqnarray}
and the general solution has the form,
\begin{eqnarray}
 a_{s}^{(H+)}(z; r_{0}) &=&
  Ae^{\alpha_i z} + B e^{\beta_i z}, \quad z > 0,
  \label{halo-eq1}\\
 a_{s}^{(H-)}(z; r_{0}) &=&
  A^{\prime}e^{\alpha_i z} + B^{\prime}e^{\beta_i z},
  \quad z < 0, 
  \label{halo-eq2}
\end{eqnarray}
where $A, A^{\prime}, B$ and $B^{\prime}$ are the integration 
constants and $\alpha_i$ and $\beta_i$ are the solutions of 
the following equation that is associated with Eq. (\ref{ahalo}),
\begin{eqnarray}
 k^2 - \frac{2 d_{A0}}{R d_{\bot}}k 
  - \left[
    \left(\frac{\nu_{s}}{R}\right)^2
    + \frac{n_{h}v\sigma_{i}}{d_{\bot}}
  \right] = 0.
  \label{ax-eq1}
\end{eqnarray}
Then the following two solutions are obtained:
\begin{eqnarray}
 \alpha_i &=& \frac{d_{A0}}{R d_{\bot}}
  -\sqrt{\left(\frac{d_{A0}}{R d_{\bot}}\right)^2 
  + \left(\frac{\nu_{s}}{R}\right)^2
    + \frac{n_{h}v\sigma_{i}}{d_{\bot}}},\\
 \beta_i &=& \frac{d_{A0}}{R d_{\bot}}
   + \sqrt{\left(\frac{d_{A0}}{R d_{\bot}}\right)^2 
  + \left(\frac{\nu_{s}}{R}\right)^2
    + \frac{n_{h}v\sigma_{i}}{d_{\bot}}}.
\end{eqnarray}
From the boundary condition in Eq. (\ref{boundary1}), Eqs. 
(\ref{halo-eq1}) and (\ref{halo-eq2}) are rewritten as
\begin{eqnarray}
  a_{s}^{(H+)}(z ; r_{0}) &=& 
      B[-\exp\{(\beta_i - \alpha_i)h\}e^{\alpha_i z} + e^{\beta_i z}] 
      \quad z > 0 \\
  a_{s}^{(H-)}(z ; r_{0}) &=& 
      B^{\prime}[-\exp\{-(\beta_i - \alpha_i)h\}e^{\alpha_i z} 
      + e^{\beta_i z}]\quad z < 0.
\end{eqnarray}
In the galactic disk, the general solution takes the form
\begin{eqnarray}
  a_{s}^{(D+)}(z; r_{0}) &=&
      Pe^{\gamma_i z} + Qe^{\delta_i z} + C^{\prime}
      \quad z \geq 0,
      \\
 a_{s}^{(D-)}(z; r_{0}) &=&
      P^{\prime}e^{\gamma_i z} + Q^{\prime}e^{\delta_i z} + C^{\prime}
      \quad z < 0,
\end{eqnarray}
where $P, P^{\prime}, Q$ and $Q^{\prime}$ are also integration 
constants and
\begin{eqnarray}
 C^{\prime} = \frac{C}{\left(\frac{\nu_{s}}{R}\right)^2
    + \frac{n_{g}v\sigma_{i}}{D_{\bot}}}.
\end{eqnarray} 
$\gamma_i$ and $\delta_i$ are the solutions to an equation 
similar to Eq. (\ref{ax-eq1})
\begin{eqnarray}
 \gamma_i &=& \frac{D_{A0}}{R D_{\bot}}
  -\sqrt{\left(\frac{D_{A0}}{R D_{\bot}}\right)^2 
  + \left(\frac{\nu_{s}}{R}\right)^2
    + \frac{n_{g}v\sigma_{i}}{D_{\bot}}},\\
 \delta_i &=& \frac{D_{A0}}{R D_{\bot}}
   + \sqrt{\left(\frac{D_{A0}}{R D_{\bot}}\right)^2 
  + \left(\frac{\nu_{s}}{R}\right)^2
    + \frac{n_{g}v\sigma_{i}}{D_{\bot}}}.
\end{eqnarray}
We suppose the cosmic ray source distributes uniformly; then
\begin{eqnarray}
 \chi(r_0) = \theta(R_i - r_0),\quad
  0 < R_i \le R.
\end{eqnarray}
Furthermore,
\begin{eqnarray}
 \int_0^{R_i} \theta(R_i - r_0)r_0
  J_0\left(\frac{\nu_s r_0}{R}\right)dr_0 =
  \frac{RR_i}{\nu_s}J_1 \left(\frac{R_i \nu_s}{R}\right),
\end{eqnarray}
and after straightforward but bit tedious calculations,
we obtain the number density of the $i$-th cosmic ray particles 
\begin{eqnarray}
  & &N_{i}^{(H+)}(E, r, z)\nonumber\\
  &=&  
  \sum_{s = 1}^{\infty}
  \frac{
    2 g_i E^{-\gamma_{i0}} R_i 
    J_{1}\left(\frac{R_i \nu_s}{R}\right)
    J_0\left(\frac{\nu_s}{R}r\right)
    }
    {
    |J_{1}(\nu_{s})|^2 \nu_s D_{\bot}R 
    \left[
      \left(\frac{\nu_{s}}{R}\right)^2
      + \frac{n_{g}v\sigma_{i}}{D_{\bot}}
    \right]
    }\nonumber\\
 & &
  \Biggl[
   \frac{1}
   {
   \left[
    (\gamma_i - \alpha_i)\exp\{(\beta_i - \alpha_i)h\}e^{\alpha_i h_{g}}
    - (\gamma_i - \beta_i)e^{\beta_i h_{g}} 
  \right]
   }\nonumber\\
 & &\quad\times
  \Bigl(
   (\delta_i - \gamma_i)
     \Bigl\{
      \left[
       (\alpha_i + \beta_i)(\gamma_i + \delta_i) - 2 \gamma_i \delta_i 
       - (\alpha_i^2 + \beta_i^2)
     \right]
      \left[
       e^{(\delta_i - \gamma_i)h_{g}} 
       - e^{-(\delta_i - \gamma_i)h_{g}}
     \right]\nonumber \\
  & &-
  \left[
    (\alpha_i - \delta_i)(\gamma_i - \beta_i)e^{(\delta_i - \gamma_i)h_{g}}
    -(\beta_i - \delta_i)(\gamma_i - \alpha_i)e^{-(\delta_i - \gamma_i)h_{g}}
  \right]
  e^{(\beta_i - \alpha_i)(h - h_{g})} \nonumber \\
  & &-
  \left[
    (\beta_i - \delta_i)(\gamma_i - \alpha_i)e^{(\delta_i - \gamma_i)h_{g}}
    -(\alpha_i -\delta_i)(\gamma_i - \beta_i)e^{-(\delta_i - \gamma_i)h_{g}}
  \right]
  e^{-(\beta_i - \alpha_i)(h - h_{g})}
  \Bigr\}^{-1} \nonumber \\
  & &\quad\times
  \Bigl\{
  \left[
    (\alpha_i + \beta_i)\gamma_i - (\alpha_i^2 + \beta_i^2)
  \right]
  (e^{\gamma_i h_{g}} - e^{-\gamma_i h_{g}}) 
  \nonumber \\
  & &-
  \left[
    \beta_i(\gamma_i - \alpha_i)e^{\gamma_i h_{g}}
    - \alpha_i(\gamma_i - \beta_i)e^{-\gamma_i h_{g}}
  \right]
  e^{(\beta_i - \alpha_i)(h - h_{g})} \nonumber \\
  & &-
  \left[
    \alpha_i(\gamma_i - \beta_i)e^{\gamma_i h_{g}}
    -\beta_i(\gamma_i - \alpha_i)e^{-\gamma_i h_{g}}
  \right]
  e^{-(\beta_i - \alpha_i)(h - h_{g})}
  \Bigr\}e^{\delta_i h_g} - \gamma_i
 \Bigr)e^{\gamma_i h_g}
  \Biggr]\nonumber\\
  & &\quad\times 
   [-\exp\{(\beta_i - \alpha_i)h\}e^{\alpha_i z} + e^{\beta_i z}],
\end{eqnarray}
\begin{eqnarray}
  & &N_{i}^{(D+)}(E, r, z)\nonumber\\ 
  &=&
  \sum_{s = 1}^{\infty}
  \frac{
    2 g_i E^{-\gamma_{i0}} R_i 
    J_{1}\left(\frac{R_i \nu_s}{R}\right)
    J_0\left(\frac{\nu_s}{R}r\right)
    }
    {
    |J_{1}(\nu_{s})|^2 \nu_s D_{\bot}R 
    \left[
      \left(\frac{\nu_{s}}{R}\right)^2
      + \frac{n_{g}v\sigma_{i}}{D_{\bot}}
    \right]
    }\nonumber \\
 & &
  \Biggl[
  \frac{1}
  {
  \left[
   (\gamma_i - \alpha_i)\exp\{(\beta_i - \alpha_i)h\}e^{\alpha_i h_{g}}
   - (\gamma_i - \beta_i)e^{\beta_i h_{g}} 
 \right]
  }\nonumber\\
  & &\quad\times
   \Bigl(
    \left[
     (\alpha_i - \delta_i)\exp\{(\beta_i - \alpha_i)h\}e^{\alpha_i h_g}
     - (\beta_i - \delta_i)e^{\beta_i h_g}
   \right]\nonumber\\
 & &\quad\times
  \Bigl\{
   \left[
    (\alpha_i + \beta_i)(\gamma_i + \delta_i) - 2 \gamma_i \delta_i 
    - (\alpha_i^2 + \beta_i^2)
  \right]\nonumber
   \left[
    e^{(\delta_i - \gamma_i)h_{g}} 
    - e^{-(\delta_i - \gamma_i)h_{g}}
  \right]
 \nonumber \\
 & &-
  \left[
   (\alpha_i - \delta_i)(\gamma_i - \beta_i)e^{(\delta_i - \gamma_i)h_{g}}
    -(\beta_i - \delta_i)(\gamma_i - \alpha_i)e^{-(\delta_i - \gamma_i)h_{g}}
 \right]
  e^{(\beta_i - \alpha_i)(h - h_{g})} \nonumber \\
 & &-
  \left[
   (\beta_i - \delta_i)(\gamma_i - \alpha_i)e^{(\delta_i - \gamma_i)h_{g}}
   -(\alpha_i -\delta_i)(\gamma_i - \beta_i)e^{-(\delta_i - \gamma_i)h_{g}}
 \right]
   e^{-(\beta_i - \alpha_i)(h - h_{g})}
  \Bigr\}^{-1} \nonumber \\
 & &\quad\times
  \Bigl\{
   \left[
    (\alpha_i + \beta_i)\gamma_i - (\alpha_i^2 + \beta_i^2)
  \right]
   (e^{\gamma_i h_{g}} - e^{-\gamma_i h_{g}}) 
 \nonumber \\
 & &-
  \left[
   \beta_i(\gamma_i - \alpha_i)e^{\gamma_i h_{g}}
   - \alpha_i(\gamma_i - \beta_i)e^{-\gamma_i h_{g}}
 \right]
  e^{(\beta_i - \alpha_i)(h - h_{g})} \nonumber \\
 & &-
   \left[
    \alpha_i(\gamma_i - \beta_i)e^{\gamma_i h_{g}}
    -\beta_i(\gamma_i - \alpha_i)e^{-\gamma_i h_{g}}
  \right]
   e^{-(\beta_i - \alpha_i)(h - h_{g})}
  \Bigr\}e^{\delta_i h_g}\nonumber\\
 & &+
   \alpha_i\exp\{(\beta_i - \alpha_i)h\}e^{\alpha_i h_g}
   - \beta_ie^{\beta_i h_g}
 \Bigr)e^{\gamma_i z}\nonumber\\
  & &+
   \Bigl\{
   \left[
   (\alpha_i + \beta_i)(\gamma_i + \delta_i) - 2 \gamma_i \delta_i 
   - (\alpha_i^2 + \beta_i^2)
  \right]
  \left[
   e^{(\delta_i - \gamma_i)h_{g}} 
   - e^{-(\delta_i - \gamma_i)h_{g}}
  \right]
  \nonumber \\
  & &-
  \left[
   (\alpha_i - \delta_i)(\gamma_i - \beta_i)e^{(\delta_i - \gamma_i)h_{g}}
    -(\beta_i - \delta_i)(\gamma_i - \alpha_i)e^{-(\delta_i - \gamma_i)h_{g}}
  \right]
  e^{(\beta_i - \alpha_i)(h - h_{g})} \nonumber \\
  & &-
   \left[
    (\beta_i - \delta_i)(\gamma_i - \alpha_i)e^{(\delta_i - \gamma_i)h_{g}}
    -(\alpha_i -\delta_i)(\gamma_i - \beta_i)e^{-(\delta_i - \gamma_i)h_{g}}
  \right]
  e^{-(\beta_i - \alpha_i)(h - h_{g})}
  \Bigr\}^{-1} \nonumber \\
  & &\quad\times
  \Bigl\{
  \left[
   (\alpha_i + \beta_i)\gamma_i - (\alpha_i^2 + \beta_i^2)
  \right]
  (e^{\gamma_i h_{g}} - e^{-\gamma_i h_{g}}) 
  \nonumber \\
  & &-
  \left[
   \beta_i(\gamma_i - \alpha_i)e^{\gamma_i h_{g}}
   - \alpha_i(\gamma_i - \beta_i)e^{-\gamma_i h_{g}}
  \right]
  e^{(\beta_i - \alpha_i)(h - h_{g})} \nonumber \\
  & &-
  \left[
    \alpha_i(\gamma_i - \beta_i)e^{\gamma_i h_{g}}
    -\beta_i(\gamma_i - \alpha_i)e^{-\gamma_i h_{g}}
  \right]
  e^{-(\beta_i - \alpha_i)(h - h_{g})}
  \Bigr\}e^{\delta_i z} + 1
  \Biggr],
\end{eqnarray}
\begin{eqnarray}
  & &N_{i}^{(H-)}(E, r, z)\nonumber\\
  &=&  
  \sum_{s = 1}^{\infty}
  \frac{
    2 g_i E^{-\gamma_{i0}} R_i 
    J_{1}\left(\frac{R_i \nu_s}{R}\right)
    J_0\left(\frac{\nu_s}{R}r\right)
    }
    {
    |J_{1}(\nu_{s})|^2 \nu_s D_{\bot}R 
    \left[
      \left(\frac{\nu_{s}}{R}\right)^2
      + \frac{n_{g}v\sigma_{i}}{D_{\bot}}
    \right]
    }\nonumber \\
 & &
  \Biggl[
  \frac{1}
  {
  \left[
   (\gamma_i - \alpha_i)\exp\{-(\beta_i - \alpha_i)h\}e^{-\alpha_i h_{g}}
   - (\gamma_i - \beta_i)e^{-\beta_i h_{g}} 
  \right]
  }\nonumber\\
 & &\quad\times
  \Bigl(
  (\delta_i - \gamma_i)
  \Bigl\{
   \left[
    (\alpha_i + \beta_i)(\gamma_i + \delta_i) - 2 \gamma_i \delta_i 
    - (\alpha_i^2 + \beta_i^2)
  \right]
   \left[
    e^{(\delta_i - \gamma_i)h_{g}} 
    - e^{-(\delta_i - \gamma_i)h_{g}}
  \right]\nonumber \\
  & &-
  \left[
    (\alpha_i - \delta_i)(\gamma_i - \beta_i)e^{(\delta_i - \gamma_i)h_{g}}
    -(\beta_i - \delta_i)(\gamma_i - \alpha_i)e^{-(\delta_i - \gamma_i)h_{g}}
  \right]
  e^{(\beta_i - \alpha_i)(h - h_{g})} \nonumber \\
  & &-
  \left[
    (\beta_i - \delta_i)(\gamma_i - \alpha_i)e^{(\delta_i - \gamma_i)h_{g}}
    -(\alpha_i -\delta_i)(\gamma_i - \beta_i)e^{-(\delta_i - \gamma_i)h_{g}}
  \right]
  e^{-(\beta_i - \alpha_i)(h - h_{g})}
  \Bigr\}^{-1} \nonumber \\
  & &\quad\times
  \Bigl\{
  \left[
    (\alpha_i + \beta_i)\gamma_i - (\alpha_i^2 + \beta_i^2)
  \right]
  (e^{\gamma_i h_{g}} - e^{-\gamma_i h_{g}}) 
  \nonumber \\
  & &-
  \left[
    \beta_i(\gamma_i - \alpha_i)e^{\gamma_i h_{g}}
    - \alpha_i(\gamma_i - \beta_i)e^{-\gamma_i h_{g}}
  \right]
  e^{(\beta_i - \alpha_i)(h - h_{g})} \nonumber \\
  & &-
  \left[
    \alpha_i(\gamma_i - \beta_i)e^{\gamma_i h_{g}}
    -\beta_i(\gamma_i - \alpha_i)e^{-\gamma_i h_{g}}
  \right]
  e^{-(\beta_i - \alpha_i)(h - h_{g})}
  \Bigr\}e^{-\delta_i h_g} - \gamma_i
\Bigr)e^{-\gamma_i h_g}
\Biggr]\nonumber\\
 & &\quad\times
  [-\exp\{-(\beta_i - \alpha_i)h\}e^{\alpha_i z} + e^{\beta_i z}],
\end{eqnarray}
\begin{eqnarray}
  & &N_{i}^{(D-)}(E, r, z)\nonumber\\
  &=&
  \sum_{s = 1}^{\infty}
  \frac{
    2 g_i E^{-\gamma_{i0}} R_i 
    J_{1}\left(\frac{R_i \nu_s}{R}\right)
    J_0\left(\frac{\nu_s}{R}r\right)
    }
    {
    |J_{1}(\nu_{s})|^2 \nu_s D_{\bot}R 
    \left[
      \left(\frac{\nu_{s}}{R}\right)^2
      + \frac{n_{g}v\sigma_{i}}{D_{\bot}}
    \right]
    }\nonumber \\
 & &
  \Biggl[
  \frac{1}
  {
  \left[
    (\gamma_i - \alpha_i)\exp\{-(\beta_i - \alpha_i)h\}e^{-\alpha_i h_{g}}
    - (\gamma_i - \beta_i)e^{-\beta_i h_{g}} 
  \right]
  }\nonumber\\
 & &\quad\times
  \Bigl(
   \left[
    (\alpha_i - \delta_i)\exp\{-(\beta_i - \alpha_i)h\}e^{-\alpha_i h_g}
    -(\beta_i - \delta_i)e^{-\beta_i h_g}
  \right]
   \nonumber\\
 & &\quad\times
  \Bigl\{
   \left[
    (\alpha_i + \beta_i)(\gamma_i + \delta_i) - 2 \gamma_i \delta_i 
    - (\alpha_i^2 + \beta_i^2)
  \right]
  \left[
   e^{(\delta_i - \gamma_i)h_{g}} 
   - e^{-(\delta_i - \gamma_i)h_{g}}
 \right]\nonumber \\
  & &-
   \left[
    (\alpha_i - \delta_i)(\gamma_i - \beta_i)e^{(\delta_i - \gamma_i)h_{g}}
    -(\beta_i - \delta_i)(\gamma_i - \alpha_i)e^{-(\delta_i - \gamma_i)h_{g}}
  \right]
   e^{(\beta_i - \alpha_i)(h - h_{g})} \nonumber \\
 & &-
  \left[
   (\beta_i - \delta_i)(\gamma_i - \alpha_i)e^{(\delta_i - \gamma_i)h_{g}}
   -(\alpha_i -\delta_i)(\gamma_i - \beta_i)e^{-(\delta_i - \gamma_i)h_{g}}
 \right]
  e^{-(\beta_i - \alpha_i)(h - h_{g})}
\Bigr\}^{-1} \nonumber \\
 & &\quad\times
  \Bigl\{
   \left[
    (\alpha_i + \beta_i)\gamma_i - (\alpha_i^2 + \beta_i^2)
  \right]
   (e^{\gamma_i h_{g}} - e^{-\gamma_i h_{g}}) 
 \nonumber \\
 & &-
  \left[
   \beta_i(\gamma_i - \alpha_i)e^{\gamma_i h_{g}}
   - \alpha_i(\gamma_i - \beta_i)e^{-\gamma_i h_{g}}
 \right]
  e^{(\beta_i - \alpha_i)(h - h_{g})} \nonumber \\
 & &-
   \left[
    \alpha_i(\gamma_i - \beta_i)e^{\gamma_i h_{g}}
    -\beta_i(\gamma_i - \alpha_i)e^{-\gamma_i h_{g}}
  \right]
   e^{-(\beta_i - \alpha_i)(h - h_{g})}
  \Bigr\}e^{-\delta_i h_g}\nonumber\\
  & &+
    \alpha_i\exp\{-(\beta_i - \alpha_i)h\}e^{-\alpha_i h_g}
    -\beta_ie^{-\beta_i h_g}
  \Bigr)e^{\gamma_i z}\nonumber\\
 & &+
  \Bigl\{
   \left[
   (\alpha_i + \beta_i)(\gamma_i + \delta_i) - 2 \gamma_i \delta_i 
   - (\alpha_i^2 + \beta_i^2)
  \right]
  \left[
    e^{(\delta_i - \gamma_i)h_{g}} 
    - e^{-(\delta_i - \gamma_i)h_{g}}
  \right] 
  \nonumber \\
  & &-
  \left[
   (\alpha_i - \delta_i)(\gamma_i - \beta_i)e^{(\delta_i - \gamma_i)h_{g}}
    -(\beta_i - \delta_i)(\gamma_i - \alpha_i)e^{-(\delta_i - \gamma_i)h_{g}}
  \right]
  e^{(\beta_i - \alpha_i)(h - h_{g})} \nonumber \\
  & &-
   \left[
    (\beta_i - \delta_i)(\gamma_i - \alpha_i)e^{(\delta_i - \gamma_i)h_{g}}
    -(\alpha_i -\delta_i)(\gamma_i - \beta_i)e^{-(\delta_i - \gamma_i)h_{g}}
  \right]
    e^{-(\beta_i - \alpha_i)(h - h_{g})}
  \Bigr\}^{-1} \nonumber \\
  & &\quad\times
  \Bigl\{
  \left[
   (\alpha_i + \beta_i)\gamma_i - (\alpha_i^2 + \beta_i^2)
  \right]
  (e^{\gamma_i h_{g}} - e^{-\gamma_i h_{g}}) 
  \nonumber \\
  & &-
  \left[
   \beta_i(\gamma_i - \alpha_i)e^{\gamma_i h_{g}}
   - \alpha_i(\gamma_i - \beta_i)e^{-\gamma_i h_{g}}
  \right]
  e^{(\beta_i - \alpha_i)(h - h_{g})} \nonumber \\
  & &-
  \left[
    \alpha_i(\gamma_i - \beta_i)e^{\gamma_i h_{g}}
    -\beta_i(\gamma_i - \alpha_i)e^{-\gamma_i h_{g}}
  \right]
  e^{-(\beta_i - \alpha_i)(h - h_{g})}
  \Bigr\}e^{\delta_i z} + 1
\Biggr].
\end{eqnarray}
We note that these solution reduce to those in \cite{berezinskii}
in the absence of antisymmetric diffusion due to particle drift. 
\section{Exponential Variation of Cosmic Ray Spectrum\label{psols}}
In this section, we discuss the features of the derived solutions,
especially focusing on the question whether the diffusion model can 
reproduce the exponential variation of the cosmic ray spectrum. 
Here, we present qualitative examples and then omit the scales 
in the following figures.

In accordance with \cite{ptuskin}, we assume the diffusion 
coefficients have the following energy dependence
\begin{eqnarray}
D_{\bot} = D_{D\bot} E^{m_D},\quad
d_{\bot} = d_{H\bot} E^{m_H},\quad
D_{A0} = D_{DA0}E,\quad
d_{A0} = d_{HA0}E,
\end{eqnarray}
where the turbulence exponents in the disk and halo are denoted by 
$m_D$ and $m_H$, respectively, and for the energy spectrum at 
the source, we choose $\gamma_{i0} = 2.2$.

\begin{figure}
  \begin{center}
    \includegraphics[scale=0.6]{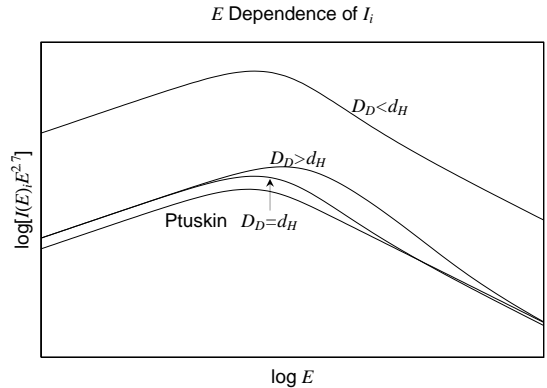}
  \end{center}
  \caption{
 Energy ($E$) Dependence of Intensity $I_i$.
 \label{edepend}}
\end{figure}

\begin{figure}
  \begin{center}
    \includegraphics[scale=0.6]{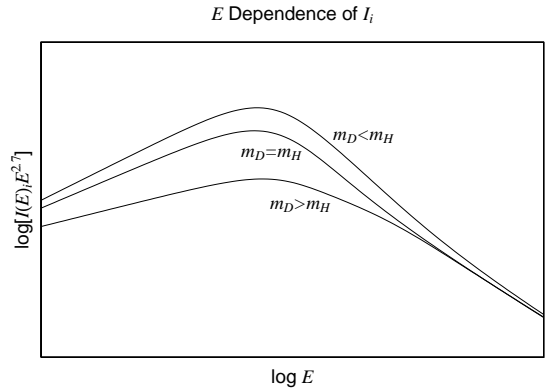}
  \end{center}
  \caption{
 $E$ Dependence of Intensity $I_i$.
 \label{mdepend}}
\end{figure}

Fig. \ref{edepend} shows the energy ($E$) dependence of the 
intensity $I_i$ of the model by \cite{ptuskin} (cited as Ptuskin) 
and out three solutions: (1) $D_{D\bot} = d_{H\bot}$ and 
$ D_{DA0} = d_{HA0}$ (shown as $D_D = d_H$); (2) the diffusion 
coefficient of the halo is larger than that of the disk, 
$D_{D\bot} < d_{H\bot}$ and $D_{D\bot} < d_{H\bot}$ 
(shown as $D_D < d_H$); and (3) the diffusion coefficient of the disk 
is larger than that of the halo,
$D_{D\bot} > d_{H\bot}$ and $D_{D\bot} > d_{H\bot}$ 
(shown as $D_D > d_H$).
In the case of our solution, the exponents of magnetic turbulence in 
the disk and the halo are equivalent, $m_D = m_H$. 
In Fig. \ref{edepend}, we multiplied $E^{2.7}$ with $I_i$ to 
emphasize the exponent variation. The three cases --- 
$D_D = d_H$, $D_D < d_H$ and $D_D > d_H$ --- can reproduce 
the variation in the cosmic ray spectrum in the same way,
as the model developed by Ptuskin.
Fig. \ref{mdepend} shows the $m$ dependence of our solutions: 
(1) the exponents of the disk and the halo are 
equivalent, $m_D = m_H$; (2) the exponent of the halo is larger 
than that of the disk, $m_D < m_H$; and (3) the exponent of 
the disk is larger than that of the halo, $m_D > m_H$. 
In all cases, we fixed $D_{D\bot} = d_{H\bot}$
and $D_{DA0} = d_{HA0}$.
We found that all the solutions exhibit a similar trend, 
as shown in \ref{edepend}. Further, we may say that the 
difference between the two exponents around the knee, 
$\Delta \gamma = \gamma_{E < E_{\rm knee}} - \gamma_{E_{\rm knee} < E}$, 
is larger value for $m_D < m_H$ than that for $m_D > m_H$.
\section{Conclusions\label{conclusion} }
We phenomenologically proposed a propagation model of
galactic cosmic rays based on the semi-empirical diffusion equation
developed by Berezinskii {\it et al.} (1990) and the diffusion tensor 
introduced by Ptuskin {\it et al.} (1993). This model takes into 
account both the symmetric diffusion and the antisymmetric 
diffusion due to the particle Hall drift. The derived solutions 
are an extension of the model developed by Ptuskin {\it et al.} 
(1993) to a two-dimensional two-layer (galactic disk and halo) model, 
and they coincide completely with the solutions derived by
Berezinskii {\it et al.} (1990) in the absence of antisymmetric 
diffusion due to particle drift. We shown that this relatively 
simple model can be used to explain the variation in the
exponent of the cosmic ray energy spectrum, $\gamma$, around 
the knee $E \approx 10^{15}$ eV.

In this paper, we showed that although the diffusive cosmic 
ray propagation model can be used to explain the observed cosmic 
ray spectrum, especially the exponential variation around the knee, 
our model is actually a more simple toy model based on assumptions 
such as the cylindrical structure of the galaxy and
simplification of magnetic field. To further test the validity of 
the diffusion model, we must conduct numerical simulations 
under more realistic situations. This is a difficult task;
nonetheless, it may help us to gain a deeper understanding of 
astroparticle physics and the galactic structure through 
which the cosmic rays pass.

\end{document}